# Bridging Mid-IR and Terahertz Domains in a Single High-Resolution Dual-Comb Spectroscopy Measurement

D. Konnov, A. Muraviev, K.L. Vodopyanov.

CREOL, The College of Optics and Photonics, University of Central Florida, Orlando, FL, USA

* Corresponding author: K.L. Vodopyanov, vodopyanov@creol.ucf.edu

**Abstract**

Dual-comb spectroscopy (DCS) is a state-of-the-art technique that utilizes a pair of broadband mutually coherent laser frequency combs to enable high-resolution, high-accuracy spectroscopic measurements with atomic-clock-level frequency referencing, and rapid, multiplexed acquisition without moving parts. DCS implementations have traditionally been confined to specific and separate spectral domains—terahertz, infrared, visible, and ultraviolet—each requiring distinct comb sources and detection mechanisms tailored to the nature of the spectroscopic target, such as molecular rotational, ro-vibrational, or electronic transitions. Yet, similar techniques may be implemented in the terahertz (THz) and mid-infrared (MIR) regions, such as optical rectification for comb generation and electro-optic sampling for detection, both utilizing crystals with $\chi^2$ nonlinearity. However, in the Reststrahlen band near the transverse optical (TO) phonon resonance in these crystals, typically between 5 and 10 THz, both linear and nonlinear susceptibilities experience abnormally high dispersion, and light propagation is strongly suppressed. This imposes a fundamental limitation, confining DCS operation to spectral regions either below or above the Reststrahlen band and effectively separating the THz and MIR domains. Here, we demonstrate high resolution DCS performed simultaneously over two broad spectral bands, each spanning an octave or higher, in the MIR (350–1150 cm$^{-1}$; 8.7–28.5 μm; 10.5–34.5 THz) and the THz range (80–160 cm$^{-1}$; 62.5–125 μm; 2.4–4.8 THz), effectively bridging these traditionally separated regions within a single measurement. This capability enables direct cross-referencing of molecular absorption line strengths across widely separated spectral domains. As a proof of concept, we demonstrate the simultaneous acquisition of ro-vibrational and pure rotational absorption spectra of ammonia (NH$_3$) with a spectral resolution of 7.3 MHz (0.00024 cm$^{-1}$), sufficient to fully resolve Doppler-broadened lineshapes across the entire measured spectral range.

**Introduction**

Laser-based dual-comb spectroscopy (DCS) is an advanced form of Fourier transform spectroscopy that exploits high temporal coherence of optical frequency combs and enables simultaneously broadband, rapid, and high-resolution spectral measurements—without the need for any moving parts [1,2,3]. Here, a frequency comb with pulse repetition rate $f_{rep}$ is used to probe the sample. The transmitted comb then interferes with a second comb, offset by a small difference $\Delta f_{rep}$ in repetition rate. The interferogram, repeating with a period of $1/\Delta f_{rep}$, is recorded over time and then Fourier-transformed to extract the sample's absorption spectrum with high spectral resolution and frequency accuracy. Over the past two decades, high-resolution DCS has rapidly advanced, especially in the mid-infrared (MIR) [4,5,6,7,8] and terahertz (THz) [9,10,11,12,13] spectral regions—key areas for molecular spectroscopy and ultrasensitive trace gas detection.

Electro-optic sampling (EOS) is a well-established technique for characterizing MIR and THz transients down to the single-cycle regime. It enables retrieval of both the electric field amplitude and phase by exploiting the transient birefringence induced in an electro-optic (EO) crystal by the incident field. This birefringence is probed using a time-delayed near-infrared (NIR) gating pulse, with the resulting polarization modulation analyzed via ellipsometry employing a balanced NIR photodetector [14,15,16].

Recent advancements employing gating pulses with down to 5-fs duration have made EOS an exceptionally sensitive technique for detecting electric fields across a broad spectral range—from the THz to the visible [17].

In the THz domain, Bartels et al. [9] were the first to demonstrate EOS-DCS with a comb span of 3–100 cm$^{-1}$ (0.1–3 THz). By eliminating the need for a mechanical delay line, they achieved a scan rate of 9 kHz while maintaining high spectral resolution (1 GHz). In their setup, one femtosecond laser was used to drive a large-area GaAs-based THz emitter, while the second laser enabled EO detection of the THz field. Finneran et al. [11] significantly improved the resolution of a THz EOS-DCS system to approximately 2 MHz by using a photoconductive antenna emitter to produce a comb spanning 5–80 cm$^{-1}$ (0.15–2.4 THz) and EO for detection of the emitted field, and measured narrow (primarily Doppler broadened) rotational transitions of water vapor at a pressure of 10 mTorr.

In the MIR, Kowligy et al. [8] used the EOS-DCS technology to achieve the spectral coverage from 370 to 3333 cm$^{-1}$ (3–27 µm, 11.1-100 THz) at a resolution of 100 MHz. To cover this frequency range, the authors used optical rectification of femtosecond NIR pulses using three different nonlinear crystals. Most recently, Konnov et al. [18] employed an EOS-based DCS system to acquire a large volume of spectral information – 200,000 comb-tooth-resolved data points with 80-MHz spacing over a 530 cm$^{-1}$ (18 THz) band in the longwave MIR range, all captured at a video-rate of 69 Hz. Overall, the key advantage of EOS is its ability to eliminate the need for cryogenically cooled photon detectors, thus enabling extremely broad spectral coverage with an added advantage of higher signal-to-noise ratio (SNR) and enhanced dynamic range.

Here, we demonstrate comb-line-resolved EOS-enabled DCS covering an exceptionally broad spectral range from 80 to 1150 cm$^{-1}$ (8.7–125 µm; 2.4–34.5 THz), excluding the Reststrahlen band between 160 and 350 cm$^{-1}$ (4.8–10.5 THz), effectively bridging the traditionally separated MIR and THz spectral regions within a single coherent measurement. As a proof of concept, we performed simultaneous acquisition of ro-vibrational and pure rotational absorption spectra of ammonia (NH$_3$), achieving a spectral resolution of 7.3 MHz (0.00024 cm$^{-1}$) and SNR of 1,000 for the strongest peaks through spectral interleaving of comb-line-resolved DCS data with progressively shifted combs over the whole MIR–THz span.

1. **Experimental setup**

As a driving source, we used a pair of mutually-coherent low-noise frequency combs based on mode-locked Cr:ZnS lasers with center wavelength λ=2.35 µm and 80-MHz comb-line spacing. Both laser combs were frequency and phase stabilized through (i) phase-locking their carrier-envelope offset, and (ii) optical locking one of the comb teeth of the second harmonic of both lasers to a common narrow-linewidth reference laser (see Methods).

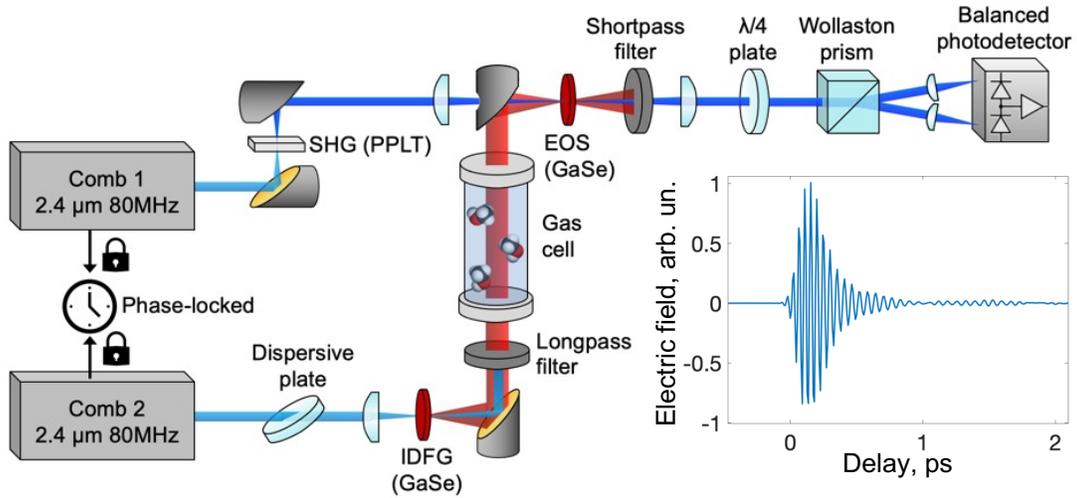

Fig. 1. Schematic of the EOS-DCS setup. A pair of phase-locked Cr:ZnS laser combs with central wavelength 2.35 µm are used as a driving source. The output of one of the Cr:ZnS combs is frequency downconverted via IDFG in GaSe to produce a broadband MIR-THz comb, while the output of the second Cr:ZnS comb is converted to NIR via second harmonic generation (SHG) and used as a source of EOS gate pulses. The inset shows a typical measured electric field waveform.

The output of one of the combs (Fig. 1) was frequency downconverted via optical rectification (also referred to as intrapulse difference frequency generation, IDFG) in a 1.3-mm thick GaSe nonlinear optical crystal to produce an ultra-broadband MIR-THz carrier-envelope-phase (CEP) stable comb that senses a molecular gas. The second comb was frequency-doubled in a 0.5-mm-thick periodically poled lithium tantalate (PPLT) crystal to generate a NIR comb used as the source of EO gate pulses. A 150-µm-thick gallium selenide (GaSe) crystal was then used for EO sampling. A gas cell (Fig. 1) was placed in the MIR-THz channel for spectral measurements.

## 2. Phase matching

In the frequency domain, EOS is understood as phase modulation of the NIR gate pulse induced by the co-propagating MIR-THz field in the EO crystal. This interaction gives rise to phase-coherent sidebands via sum- and difference-frequency generation (SFG and DFG) processes [16,17]. As a result, a multiheterodyne mixing takes place in the NIR domain, where the SFG or DFG comb (or both) interferes with the NIR gate comb, which serves as a local oscillator. This interference produces a radio-frequency comb that maps the MIR-THz comb spectrum downscaled by $f_{rep}/\Delta f_{rep}$.

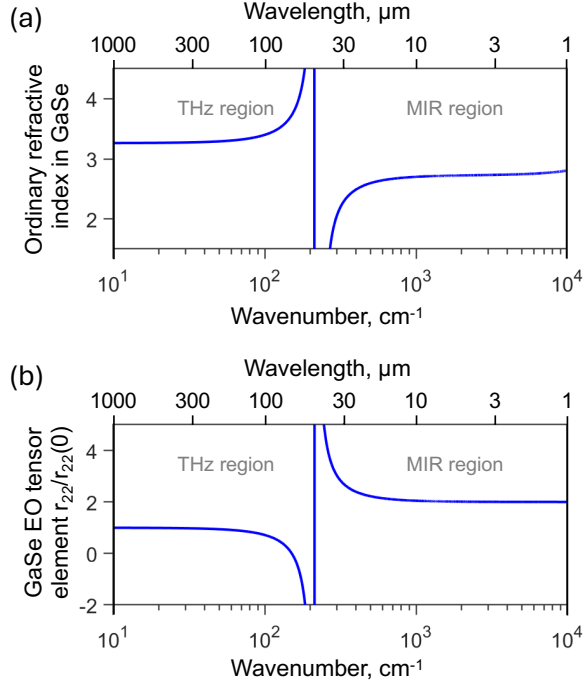

Fig. 2. (a) Ordinary refractive index ($n_o$) and (b) the normalized EO coefficient ($r_{22}$) of GaSe plotted as a function of frequency.

Both IDFG and EOS are second-order nonlinear optical processes – one being the inverse of the other – and require phase matching. In GaSe crystal that we used in the experiment, phase matching is achieved by proper selecting the polarizations of the interacting waves and adjusting the polar (θ) angle of the crystal. Fig. 2 shows the frequency dependence of the ordinary refractive index (based on data from [19]) and the electro-optic (EO) tensor component (data from [20]) for GaSe, highlighting significant dispersion in both quantities in the Reststrahlen band. Despite these anomalies, simultaneous phase matching in GaSe appears to be feasible for both the MIR and THz regions, enabling generation via IDFG and detection via EOS.

Figure 3a illustrates the orientation of a z-cut GaSe crystal used for the IDFG process. The 2.35 μm pump pulse is polarized at approximately 45° relative to the crystal's x-axis, thus providing both ordinary (*o*-wave) and extraordinary (*e*-wave) components. For MIR generation, phase matching is achieved when the blue spectral component of the pump is an *e*-wave, the red component as an *o*-wave, and the generated MIR output is also an *o*-wave. This corresponds to the *ooe* phase matching (the indices show the polarization states of the interacting waves in order of increasing frequency). For THz generation, the roles of the two pump components are reversed: the blue component is now an *o*-wave, and the red component is an *e*-wave, again producing an *o*-wave output (*oeo* phase matching). Thus, both the MIR and THz outputs share the same polarization state, preserving coherence across the combined spectral bandwidth and allowing downstream polarization-sensitive detection. On the quantum level, the IDFG phase matching process can be expressed as follows:

$$\omega_{MIR}^{o-wave} = \omega_{pumpB}^{e-wave} - \omega_{pumpR}^{o-wave} \quad (1)$$

$$\omega_{THz}^{o-wave} = \omega_{pumpB}^{o-wave} - \omega_{pumpR}^{e-wave} \quad (2)$$

Here ω is the photon frequency and "*pumpB*" and "*pumpR*" correspond to the blue and red wings of the driving laser spectrum, respectively.

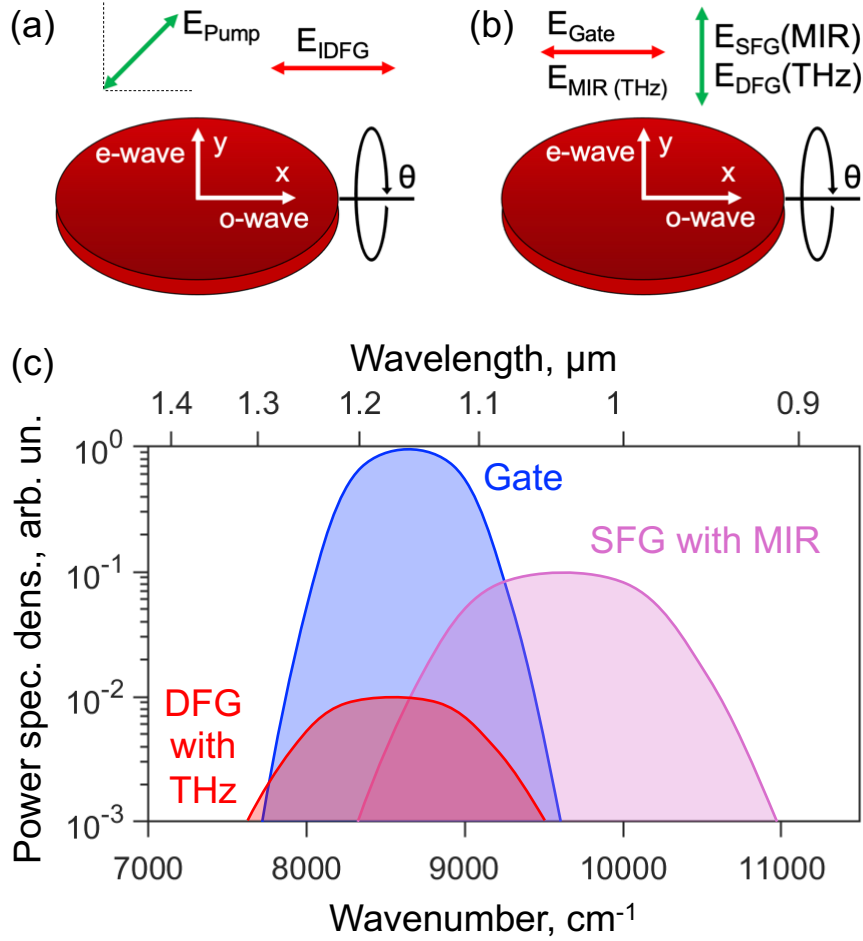

Fig. 3. (a) Orientation of the z-cut GaSe crystal for IDFG, as seen by the incoming beams. (b) Orientation of the z-cut GaSe crystal for EOS. (c) A qualitative illustration of the gate spectrum, as well as MIR and THz spectra upconverted to the NIR domain.

For EOS, we used a GaSe crystal with a similar orientation (Fig. 3b). The incoming beams—the 1.15 µm gate pulse and the MIR-THz input fields—were polarized along the ordinary axis (x-axis) of the crystal and were coherently upconverted into the orthogonal (extraordinary) polarization through the phase-matched SFG and DFG processes, correspondingly:

$$\omega_{SFG}^{e-wave} = \omega_{gate}^{o-wave} + \omega_{MIR}^{o-wave} \quad (3)$$

$$\omega_{DFG}^{e-wave} = \omega_{gate}^{o-wave} - \omega_{THz}^{o-wave} \quad (4)$$

As in the IDFG case, the phase matching type was *ooe* for the MIR, and *oeo* for THz. A qualitative illustration of the spectra for the gate and upconverted MIR-THz waveforms is shown in Fig. 3c.

Fig. 4(a,b) shows color-coded 3D simulated plots of the phase-matching function, $sinc^2(\Delta kL/2)$, for IDFG in GaSe vs. the polar angle (θ) and frequency for the generated THz and MIR components, based on the Sellmeier data [19]. Here $\Delta k$ is the wave-vector mismatch, and $L$ is the crystal length equal to 1.3 mm in our simulations. These plots indicate that simultaneously phase matched IDFG can be achieved in both spectral regions at $\theta_0 \approx 11.5°$. The corresponding 2D phase-matching plots at a fixed angle $\theta_0$ are shown in Fig. 4(c,d). The dashed lines represent the simulation for a thinner ($L$=650 µm) GaSe crystal.

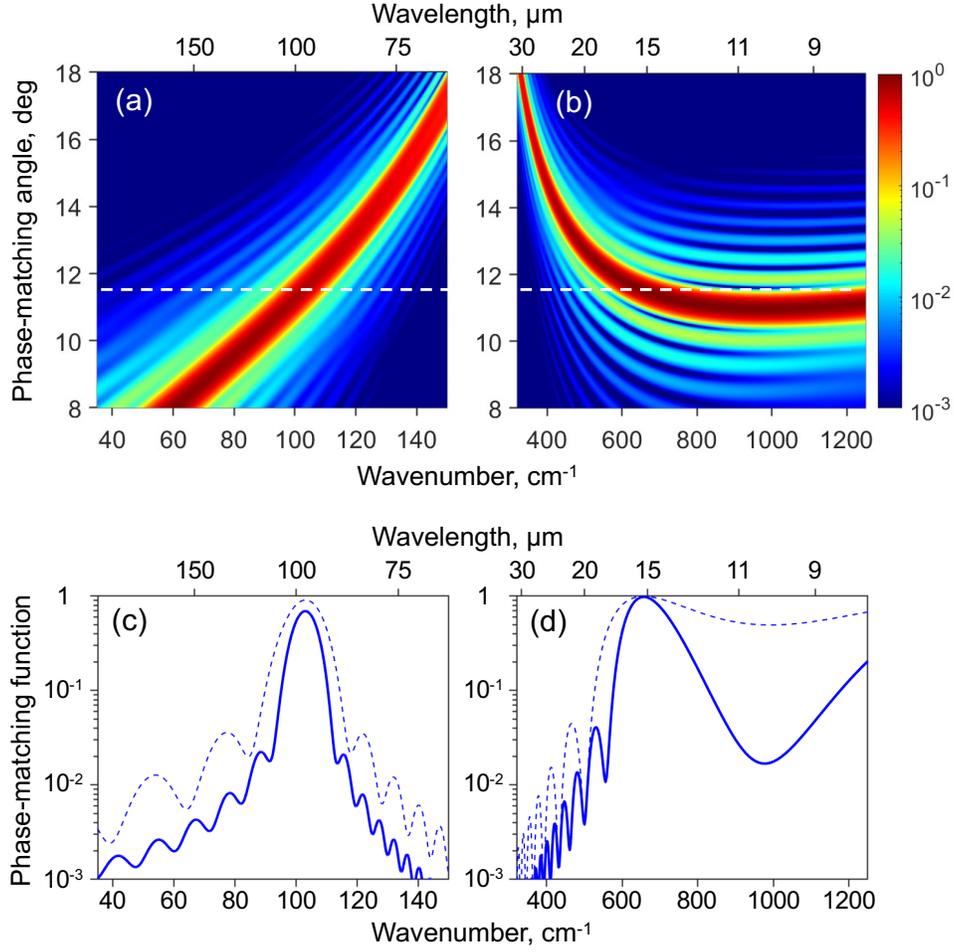

Fig. 4. (a,b) Color-coded 3D phase-matching function $sinc^2(\Delta kL/2)$ for IDFG in an $L$=1.3 mm GaSe for (a) THz and (b) MIR regions. The dashed lines in (a,b) correspond to the optimal phase-matching angle. (c,d) Cross-sections of the plots (a,b) at a fixed $\theta \approx 11.5°$. The dashed lines in (c,d) correspond to a thinner, $L$=650 μm GaSe.

Similarly, Fig. 5(a,b) shows 3D color-coded plots for the EOS phase-matching function, $sinc^2(\Delta kL/2)$, for the gate pulse central wavelength of λ=1.15 μm and an $L$=150-μm-thick GaSe. Here, the phase matched EOS detection can be achieved in both spectral regions at $\theta_0 \approx 9.5°$. The corresponding 2D plots at a fixed angle $\theta_0$ are shown in Fig. 5(c,d). One can see that while the phase-matching function is maximized in the THz region, in the MIR it remains below unity. The dashed lines in Fig. 5(c,d) represent the phase-matching functions for a thinner crystal ($L$ = 75 μm), showing a noticeable improvement in upconversion phase-matching function, particularly in the MIR. However, the crystal with this thickness was not available during our experiment.

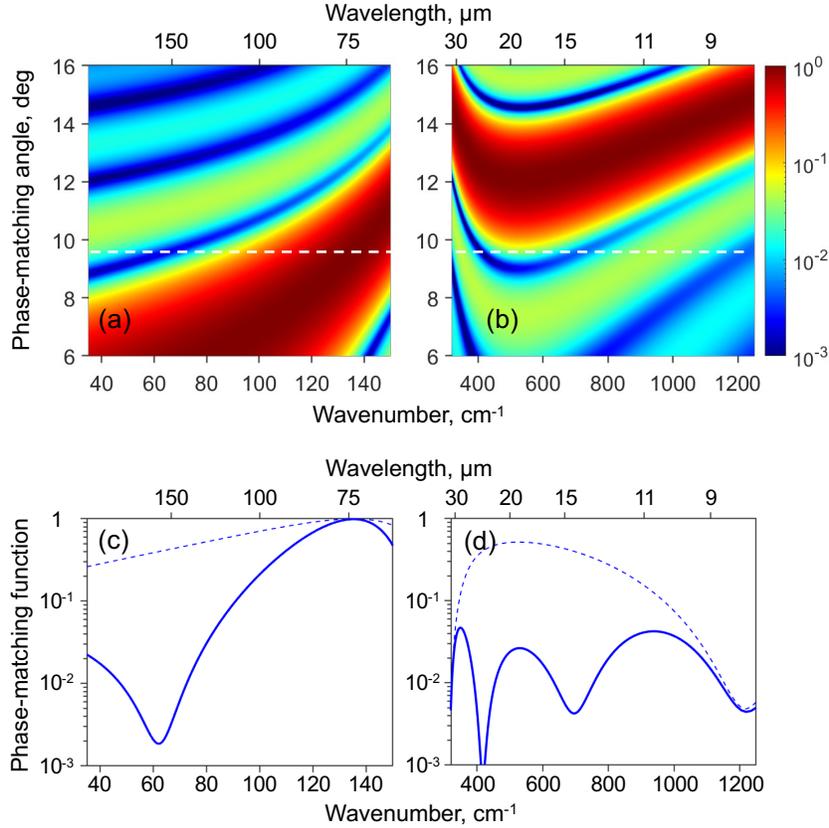

Fig. 5. (a,b) Color-coded 3D phase-matching function $sinc^2(\Delta kL/2)$ for upconversion in an $L$=150-μm-thick GaSe crystal for (a) THz and (b) MIR regions. The dashed lines in (a,b) correspond to the optimal phase-matching angle (c,d) Cross-sections of the plots (a,b) taken at a fixed θ≈9.5º. The dashed lines in (c,d) correspond to a thinner, $L$=75 μm GaSe.

Fig. 6(a) shows the produced IDFG frequency comb, measured using EOS-DCS. The comb spans two distinct, mutually coherent broadband spectral regions: a MIR band (350–1150 cm$^{-1}$; 8.7–28.5 μm; 10.5–34.5 THz) and a THz band (80–160 cm$^{-1}$; 62.5–125 μm; 2.4–4.8 THz), separated by the Reststrahlen band. The measured spectral coverage is in excellent agreement with the simulated phase-matching functions shown in Figs. 4–5. The bandwidth of the THz region was primarily limited by phase-matching, while in the MIR region, the short-wavelength cutoff was constrained by a limited spectral bandwidth of the gate pulse, which was generated directly from the laser without additional spectral broadening and temporal compression. Within the THz region, absorption features due to atmospheric water vapor are clearly visible. In the MIR region, a small narrow absorption feature at 667 cm$^{-1}$ corresponds to carbon dioxide in ambient air. A sharp intensity drop at 355 cm$^{-1}$ is caused by lattice absorption in the germanium substrate of the longpass filter (Fig. 1). Additionally, strong modulation of the spectral envelope between 355 and 550 cm$^{-1}$ arises from oscillations in the phase-matching function, in agreement with the pot in Fig. 4(d).

## 3. Measurements of ammonia

For our pilot demonstration of high-resolution spectroscopy simultaneously conducted in the MIR and THz spectral regions, we selected the ammonia (NH$_3$) molecule [21]. Fig. 6(b) displays the simulated (HITRAN database [22]) absorption cross-section spectrum of low-pressure ammonia (NH$_3$) that has distinct absorption bands around 100 cm$^{-1}$ and 950 cm$^{-1}$ corresponding to purely rotational and ro-vibrational transitions, respectively. One can see from Fig. 6 that the spectrum of our frequency comb overlaps well with these two absorption bands. The measurements were conducted using a 45-cm-long gas cell with 0.5-mm-thick CVD diamond windows at the Brewster angle. The cell was filled with ammonia at a concentration between 0.1% and 1%, using air as the buffer gas. The spectra were measured at two different pressures: 15 mbar and 2 mbar.

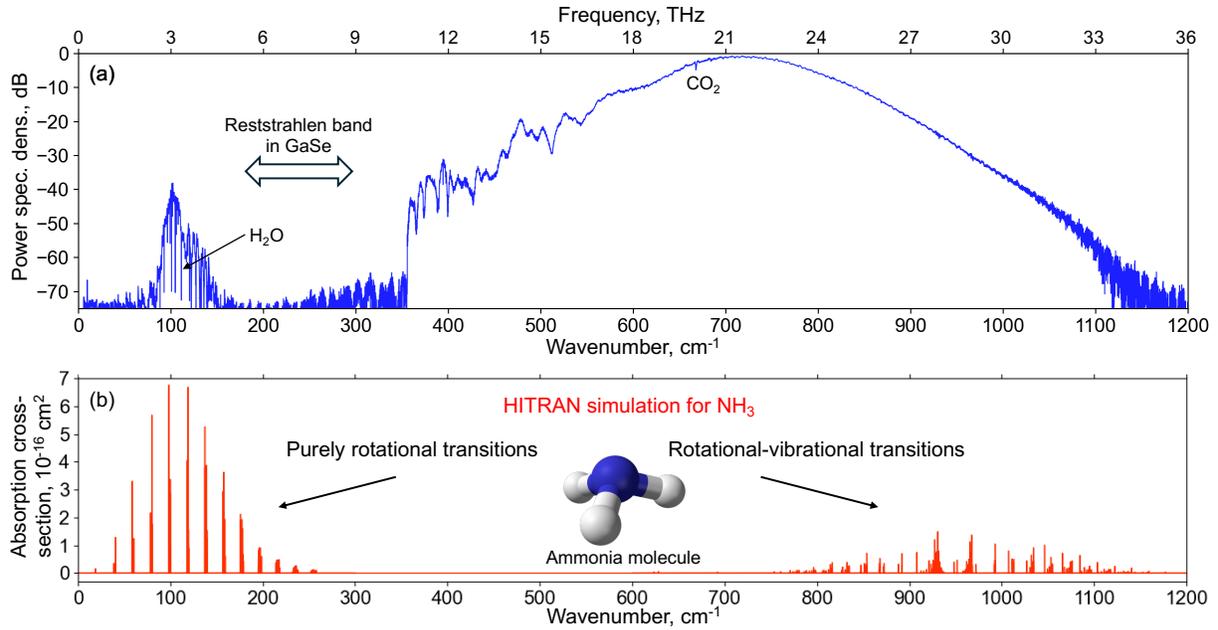

Fig. 6. (a) The MIR-THz frequency comb spectrum. (b) Simulated absorption spectrum of $NH_3$ molecule in the 10-1200 $cm^{-1}$ range (1% $NH_3$ in air, total pressure 2 mbar).

*Measurements at 15 mbar pressure.*

At a total (buffer gas plus ammonia) pressure $p$=15 mbar, the absorption linewidths were dominated by pressure broadening and were on the order of 100 MHz for both THz and MIR absorption bands. We used the data point spacing equal to the spacing of our resolved comb modes (80 MHz), which was sufficient to acquire an overview spectrum of ammonia. Portions of the EOS-DCS absorbance spectrum for the MIR and THz regions are shown in Figs. 7 and 8, respectively. (The absorbance in these plots is defined as $A$=-ln($I/I_0$), where $I$ is the transmitted intensity with the filled cell, and $I_0$ is the intensity with an empty cell.) In both spectral regions, we observed excellent agreement between the measured spectra and HITRAN simulation for ammonia [22] shown as downward-facing peaks. In the THz range (Fig. 8), one can also see absorption lines from water vapor present in the gas cell. These lines also show good agreement with the HITRAN simulation shown in gray color.

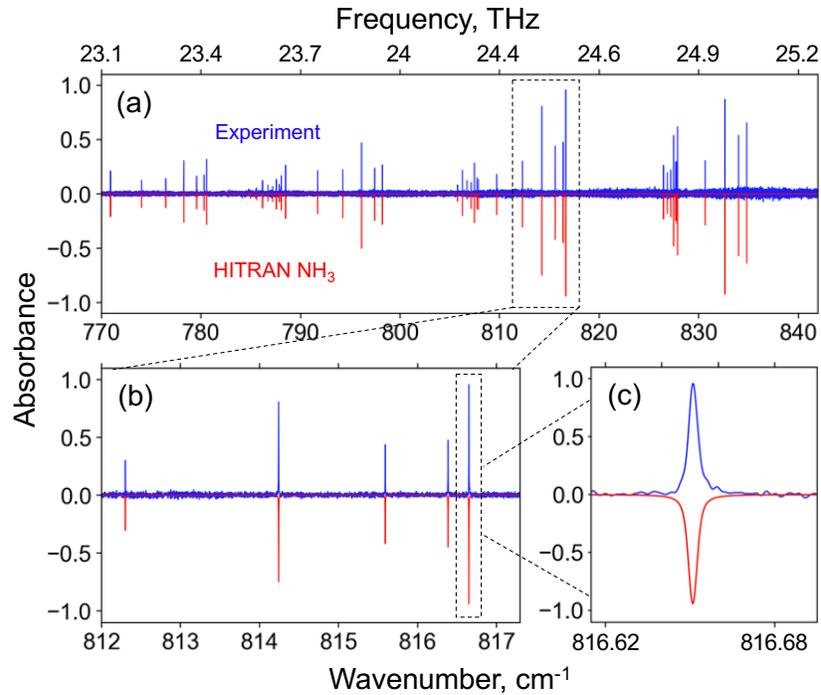

Fig. 7. High-resolution spectrum of ammonia at $p$=15 mbar overall pressure in the MIR region. HITRAN simulations are shown as downward-facing peaks. The *Absorbance* is defined as $A=-\ln(I/I_0)$, where $I$ is the transmitted intensity with the filled cell, and $I_0$ is the intensity with an empty cell.

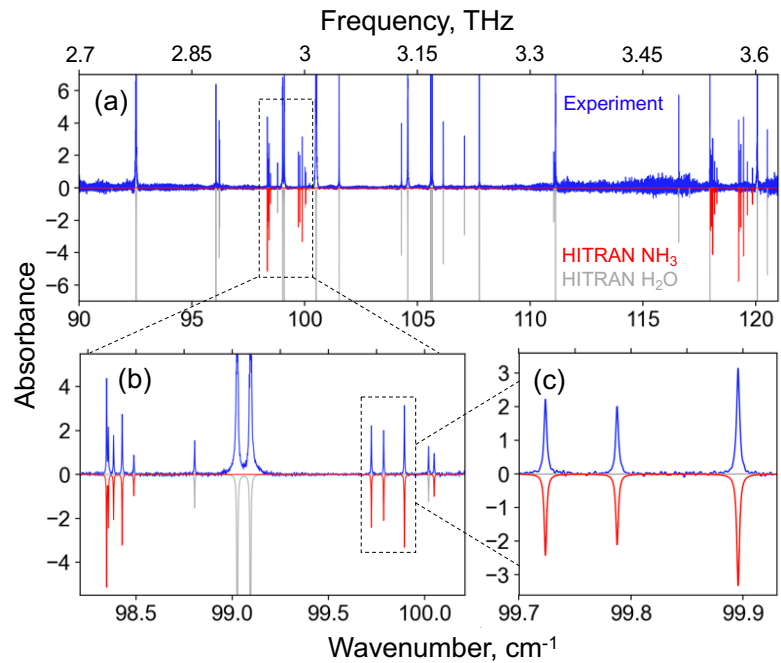

Fig. 8. High-resolution spectrum of ammonia at $p$=15 mbar overall pressure in the THz region. HITRAN simulations for ammonia and water vapor are shown as downward-facing peaks in red and gray color, correspondingly.

*Measurements at 2 mbar pressure.*

To demonstrate the high-resolution capability of our EOS-DCS method, we reduced the gas pressure to 2 mbar, thereby minimizing pressure broadening and yielding predominantly Doppler-broadened linewidths. Fig. 9 shows portions of the obtained high-resolution EOS-DCS absorbance spectrum of ammonia including MIR and THz bands. The spectrum was acquired by interleaving 11 comb-mode-resolved spectra with gradually shifted comb lines resulting in the average sampling interval of 80/11=7.3 MHz, similar to [18].

This enabled the resolution of absorption lines with full widths at half maximum (FWHM) of 21 MHz in the THz region, in agreement with the HITRAN-simulated value of 22 MHz (Fig. 5b), and 80 MHz in the MIR region, compared to the simulated value of 81 MHz (Fig. 5d). For the strongest absorption peaks in the THz spectrum, the achieved signal-to-noise ratio (SNR) was close to 1,000, where SNR is defined as the ratio of the peak height to the standard deviation of the baseline.

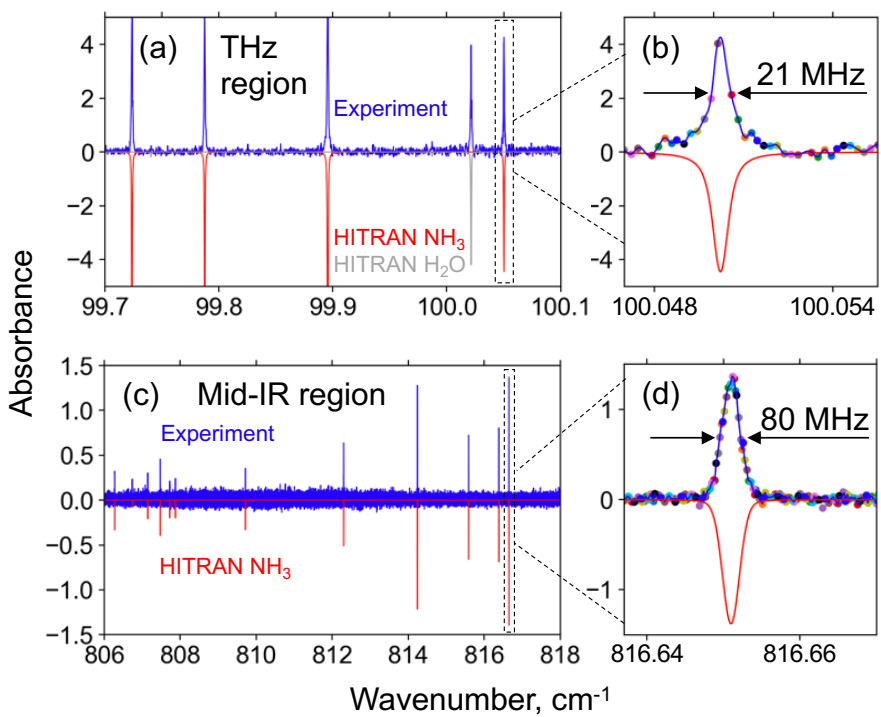

Fig. 9. High-resolution spectrum of ammonia at *p*=2 mbar total pressure; (a,b) in the THz and (c,d) in the MIR region. The spectrum was acquired by interleaving 11 gradually shifted comb-mode-resolved spectra. The experimental points in (b) and (d) are color-coded to represent measurements with shifted comb frequencies. HITRAN simulations for ammonia and water vapor are shown as downward-facing peaks in red and gray color, correspondingly.

## 4. Conclusion

We have extended high-resolution dual-comb spectroscopy to a record-wide spectral range of 6.6–200 μm, producing 1/3 of a million spectral data points effectively bridging the MIR and THz regions in a single measurement (without spectral interleaving). As a proof of concept, we simultaneously recorded ro-vibrational and pure rotational spectra of ammonia with a spectral resolution of 7.3 MHz (0.00024 cm$^{-1}$), resolving predominantly Doppler-limited linewidths in both spectral regions. With sub-kHz comb-tooth linewidths and the use of interleaving techniques, our approach enables spectral resolution on the comparable scale, with absolute frequency accuracy traceable to an atomic clock. This capability paves the way for direct, high-precision referencing of absorption line strengths between rotational and ro-vibrational transitions, and potentially other molecular modes such as internal rotation. Looking ahead, this platform opens promising avenues for the detection of trace molecular conformers with permanent dipole

moments—for instance, identifying tunneling transitions between the cis- and gauche- forms of 1,3-butadiene [23] and also for measuring the inter- and intra- molecular modes of hydrogen bonded clusters in a single experiment. Moreover, it lays the groundwork for real-time, noninvasive medical diagnostics across the MIR–THz range, such as detecting volatile biomarkers in human breath [24]. Using thinner nonlinear crystals for both IDFG and EOS opens the potential to extend the spectral coverage from below 3 μm to as far as 600 μm (0.5–100 THz). Also, it would be particularly interesting to explore spectroscopic measurements within the Reststrahlen band using thin crystals. These measurements could benefit from a resonantly enhanced EO coefficient, despite the associated increase in absorption. Strong variations in both properties are not expected to affect the measured molecular spectra, as the results are normalized to reference measurements taken with an empty gas cell.

## 5. Methods

As a driving source, we used a pair of mutually-coherent low-noise frequency combs based on mode-locked Cr:ZnS lasers with center wavelength λ=2.35 μm, average power 3 W, pulse duration 30 fs, and repetition rate 80-MHz. The carrier-envelope offset of both lasers was stabilized, similar to [18], through 3$f$-to-4$f$ interferometry, and the optical locking – by phase locking one of the comb teeth of the second harmonic of both lasers to a common narrow-linewidth (10 Hz) 1064-nm reference laser, locked in its turn to ultra-low expansion (ULE) glass cavity (Time-Base, Germany).

The output of one of the combs (Fig. 1) was frequency downconverted via IDFG using a 1.3-mm thick GaSe nonlinear optical crystal to produce an ultra-broadband MIR-THz carrier-envelope-phase (CEP) stable comb that senses a molecular gas. Before entering GaSe, the 2.35-μm pump pulses are pre-chirped using thin (~ 1 mm) YAG or sapphire plates – to improve down-conversion efficiency and optimize the spectral shape of IDFG pulses. The second comb was frequency doubled in a 0.5-mm-thick periodically poled lithium tantalate (PPLT) crystal to produce a NIR comb serving as a source of EOS gate pulses.

The MIR-THz and NIR gate beams were combined using an off-axis parabolic 90º mirror reflecting the MIR-THz beam with a 2-mm diameter hole to transmit the NIR gate beam (Fig. 1). The combined beams were focused into the 150-μm-thick GaSe crystal for EO sampling. After the EOS crystal, the gate pulse was shortpass filtered and sent to an ellipsometry setup consisting of a quarter-wave plate, a Wollaston prism, and a balanced photodetector that uses two InGaAs photodiodes (Thorlabs, model PDB450C, bandwidth 45 MHz). An attenuation filter wheel (not shown in Fig. 1) was used to keep the total incoming NIR power for each detector below saturation. Since the upconverted MIR and THz signals overlap with different spectral regions of the NIR gate pulse, appropriate spectral filters were used to balance the detection efficiency across the two frequency ranges. Additionally, Brewster plates were employed as partial polarizers to optimize the relative intensities of the two orthogonal polarization components emerging from the EO crystal. The obtained interferograms from the balanced photodetector were digitized using a 16-bit analog-to-digital converter (AlazarTech ATS9626) and coherently averaged.

To improve signal-to-noise ratio we used coherent averaging of interferograms similar to traditional FTIR spectroscopy. At the repetition rate offset between the two combs of $\Delta f_{rep}$ = 92 Hz, we typically averaged 100,000–1,000,000 interferograms for each spectrum. The spectra were retrieved via fast Fourier transform (FFT) and normalized to the spectrum taken with the empty gas cell; in most cases, the baseline did not require any correction. All frequencies of our system, including the repetition rates of both lasers, the beat notes between the comb teeth and the narrow-linewidth reference laser, and the analog-to-digital converter clock, are referenced to a rubidium (Rb) atomic clock with a fractional uncertainty of <$10^{-10}$. This results in the absolute frequency uncertainty of <3 kHz ($10^{-7}$ cm$^{-1}$) in the MIR, and <300 Hz ($10^{-8}$ cm$^{-1}$) in the THz domain.

For the absolute absorption cross-sections, based on the known optical path length, sample concentration, temperature, and pressure, the estimated fractional uncertainty of our measurements is approximately 3–5%. However, because the broad spectrum is acquired simultaneously, the *relative* absorption cross-sections (or line strengths) across different spectral regions can be determined with a fractional accuracy as high as 0.1%.

**Acknowledgements**

This work was supported through funding from the US Office of Naval Research (ONR) award N00014-18-1-2176; US Air Force Office of Scientific Research (AFOSR) awards FA9550-23-1-0126 and FA9550-24-1-0196; and U.S. Department of Energy, award DE-SC0012704.

**Key words:** rotational spectra, ro-vibrational spectra, ammonia, absorption line strength, high resolution spectra, frequency combs, dual-comb spectroscopy